# *Orbital Angular Momentum-based Space Division Multiplexing for High-capacity Underwater Optical Communications*

Yongxiong Ren[1]*, Long Li[1]*, Zhe Wang[1], Seyedeh Mahsa Kamali[2], Ehsan Arbabi[2], Amir Arbabi[2], Zhe Zhao[1], Guodong Xie[1], Yinwen Cao[1], Nisar Ahmed[1], Yan Yan[1], Cong Liu[1], Asher J. Willner[1], Solyman Ashrafi[3], Moshe Tur[4], Andrei Faraon[2], and Alan E. Willner[1]

[1]Department of Electrical Engineering, University of Southern California, Los Angeles, CA 90089, USA.

[2]T. J. Watson Laboratory of Applied Physics, California Institute of Technology, Pasadena, CA 91125, USA.

[3]NxGen Partners, Dallas, TX75219, USA.

[4]School of Electrical Engineering, Tel Aviv University, Ramat Aviv 69978, Israel.

Corresponding email: yongxior@usc.edu, willner@usc.edu

*These authors contributed equally to this work

**Abstract:** To increase system capacity of underwater optical communications, we employ the spatial domain to simultaneously transmit multiple orthogonal spatial beams, each carrying an independent data channel. In this paper, we multiplex and transmit four green orbital angular momentum (OAM) beams through a single aperture. Moreover, we investigate the degrading effects of scattering/turbidity, water current, and thermal gradient-induced turbulence, and we find that thermal gradients cause the most distortions and turbidity causes the most loss. We show systems results using two different data generation techniques, one at 1064 nm for 10-Gbit/s/beam and one at 520 nm for 1-

**Gbit/s/beam; we use both techniques since present data-modulation technologies are faster for infrared (IR) than for green. For the higher-rate link, data is modulated in the IR, and OAM imprinting is performed in the green using a specially-designed metasurface phase mask. For the lower rates, a green laser diode is directly modulated. Finally, we show that inter-channel crosstalk induced by thermal gradients can be mitigated using multi-channel equalisation processing.**

There is growing interest in high-capacity underwater wireless communications in order to support the significant increase in the demand for data, such as from sensor networks, unmanned vehicles and submarines[1-5]. Traditionally, acoustic waves have been used for underwater communications, but this technique has quite limited bandwidth capacity[1, 2, 5-7]. Alternatively, optics–especially for the low-attenuation blue-green region–can enable higher-capacity underwater transmission links due to the much higher carrier-wave frequency[8-15]. In order to increase the capacity of underwater communications, a laudable goal would be to simultaneously transmit multiple independent data channels by using the spatial domain for multiplexing, i.e., space division multiplexing (SDM)[16]. If the beams are mutually orthogonal, the different beams can then be efficiently (de-)multiplexed, transmitted through a single transmitter/receiver aperture pair, and co-propagate with little inherent crosstalk.

An orthogonal spatial modal basis set that might enable underwater SDM is orbital angular momentum (OAM) modes[17]. A light beam with a helical wavefront carries an OAM value corresponding to $\ell\hbar$ per photon, where $\hbar$ is the reduced Planck's constant and $\ell$ is an unbounded integer that represents the number of $2\pi$ phase changes in the azimuthal direction[17, 18]. The phase front of an OAM beam twists along the propagation direction and results in a ring-shaped intensity profile with a central null[18]. Previous reports have explored the use of OAM multiplexing for high-capacity data transmission through the atmosphere using 1.55-µm light[19-23]. In general, free-space systems may need to deal with atmospheric turbulence, which can disrupt the beams' phase fronts and cause intermodal crosstalk[24-27].

Much has been uncovered in free-space OAM systems in the infrared (IR), yet little has been reported for underwater blue-green communications. Indeed, the underwater environment presents several different challenges for a high-speed OAM link[28, 29]. For example, the OAM beam itself and the data it carries can be significantly degraded due to various widely-varying effects, such as dynamic scattering/turbidity, water currents, and temperature gradients[12-15, 30-34]. Although these issues are challenging for non-OAM, single beam underwater links, the problem may escalate for systems using phase front-sensitive OAM beams[25, 33-34].

Recent reports have shown a 4.8-Gbit/s underwater link using a Gaussian beam by directly modulating a 1.2-GHz bandwidth 450-nm laser diode with orthogonal-frequency-division-multiplexing (OFDM) data[11]. Moreover, it has been shown recently that a blue OAM beam can propagate through 3-metre of water, which includes a scattering solution[28]. However, little has been reported on the performance of OAM-multiplexed underwater data transmission or its degradation due to underwater effects.

In this paper, we explore OAM multiplexing for high-speed underwater communications, and we transmit four multiplexed green OAM beams through 1.2-metre of water[35]. Furthermore, we investigate the impact of various underwater conditions (e.g., scattering/turbidity, current, and thermal gradients) on beam quality and system performance, finding that thermal gradients can produce significant beam-quality degradation (e.g., modal distortion and beam wander). Importantly, we show systems results using two different approaches for data modulation, one at 10-Gbit/s/beam in the infrared (IR) and one at 1-Gbit/s/beam in the green); we show both approaches since data modulation technologies are currently faster in the IR[5, 36]. For the IR approach, we modulate a 1064-nm beam at 10-Gbit/s/beam and frequency double it into the green by using a periodically poled lithium niobate (PPLN) nonlinear crystal, and a specially designed integrated dielectric metasurface phase mask[37] imprints the OAM on the beam; note that this 40-Gbit/s aggregate capacity is ~8 times higher than the previously reported result using a conventional Gaussian beam[11]. For the green approach, we directly modulate the 532-nm laser diode. Finally, in order to take advantage of the multiple beams traversing the same medium, we demonstrate that inter-channel crosstalk induced by thermal gradients can be mitigated using a multi-channel equalisation digital signal processing (DSP) algorithm at the receiver[38].

**Results**

Figure 1 illustrates a prospective application scenario of using OAM multiplexing for high-speed underwater data transmission. We explore such a scenario under laboratory conditions to help determine the challenges of OAM-based SDM underwater communications.

**OAM beam propagation through various water conditions**

We first investigate the influence of underwater propagation on green OAM beams. In general, a light beam propagating through water may suffer degradation from various effects, including scattering/turbidity, currents, and turbulence. We emulate these underwater conditions in a 1.2-metre-long rectangular tank (with 17 cm in width and 30 cm in height) filled with tap water. Specifically, underwater scattering/turbidity is produced with suspensions of $Al(OH)_3$ and $Mg(OH)_2$, which are obtained by adding a commercial antacid preparation (Maalox®)[8, 12, 14]. Circulation pumps pointing perpendicular to the propagation direction are evenly placed along the link path inside the water tank to produce a water current. Additionally, thermal gradient-induced water turbulence is created by introducing temperature inhomogeneity along the optical link, which is accomplished via mixing room temperature and heated water. The measured power loss induced by traversing the tank and the 1.2 metres of tap water is around 2.5 dB, which is mainly caused by the reflections at the tank's glass interfaces, with the power loss incurred by the water itself being negligible.

Figure 2(a) shows the intensity profiles of the individually transmitted and received Gaussian ($\ell =0$) and OAM beams ($\ell =+1$ and $+3$) at 520 nm under various conditions: (i) with only tap water (a1-a3), (ii) with water current (a4-a6), (iii) with the Maalox solution (a7-a9), and (iv) with a thermal gradient (a10-a12). The OAM beam with either $\ell =+1$ or $+3$ is generated by shining the Gaussian beam onto a spatial light modulator (SLM) loaded with a helical phase pattern of $\ell =+1$ or $\ell =+3$. The water current in Figs. 2(a4-a6) is created using three circulation pumps, each with a flow rate of 26.8 litres per min. The Maalox solution added into the water tank containing 30 litters of water is 1.5-millilitre of 0.5% diluted Maalox, and the water after adding the Maalox is circulated by pumps for 1 minute to obtain a uniform scattering suspension (see Fig. S1(a-b) in Supplementary Section 1 for the case of a nonuniform suspension when there is no added circulation). The room temperature and heated water that are mixed for turbulence emulation have a temperature difference of 0.2 $^{o}$C; such an approach has been used previously to emulate thermal gradients in water[33, 34].

We see that the ring-shaped intensity profiles of the OAM beams tend to be maintained after propagating through tap water, and are slightly distorted by the water current. When a 1.5-millilitre Maalox solution is added into still water, there is a small, time-varying change in the intensity profiles that might be a result of the natural dynamic diffusive movement of Al(OH)$_3$/Mg(OH)$_2$ particles in the water. When the particles become evenly distributed in the water after 1-minute circulation from one pump, the distortions of the OAM intensity profiles tend to be small (Figs. 2(a7-a9)). However, an additional power loss of 4.5 dB to the link is measured. We expect a larger power loss for a higher concentration of scattering particles.

Figures 2(a10-a12) depict snapshots of intensity profiles under thermal gradient-induced turbulence, showing significant distortions in the beam profiles. We believe that this is mainly due to the higher-order wavefront aberrations that can result from the refractive index inhomogeneity induced by the water thermal gradient. Moreover, the thermal gradient introduces a dynamic beam wander at the receiver, as depicted in Fig. 2(b). The maximal displacement of the received beams is estimated to be ~1 mm, which is expected to increase under a larger thermal gradient (as shown in Supplementary Fig. S1(c)). For comparison, the statistic for the beam wander due to water current is also shown. The beam wander combined with other higher-order wavefront aberrations could cause the spreading of the transmitted OAM beam power into neighbouring modes, resulting in significant performance degradation (Supplementary Fig. S1(d)). Figure 2(c) shows the OAM power spectrum for beam $\ell$ =+3 under the above underwater conditions. The crosstalk values onto adjacent modes increase by 0.5 and >7 dB with current and turbulence, respectively. Figure 2(d) presents the power transfer between OAM modes $\ell$ =±1 and ±3 under water current. It is estimated that the total crosstalk for each mode is below -10.3 dB if all four beams are simultaneously transmitted.

Given the above measurements, it seems that the thermal gradient-induced turbulence has a larger impact on beam quality than scattering or current, yet the Maalox-induced scattering (if uniformly distributed) may introduce significant link loss.

**System performance measurements of four OAM multiplexed underwater links**

In this section, we present the system performance measurements when simultaneously transmitting four OAM beams. Two OAM multiplexed underwater links each using a different source data generation technique are demonstrated. The first link transmits a 1-Gbit/s signal at 520 nm on each beam using the direct modulation of a laser diode, resulting in a capacity of 4 Gbit/s. For the second link, each beam carries a 10-Gbit/s signal generated using frequency doubling of a data signal at 1064 nm, achieving a significantly higher capacity of 40 Gbit/s.

**4-Gbit/s data link using directly modulated laser diodes:** Two 1-Gbit/s on-off-keyed (OOK) signal beams at 520 nm are generated by directly modulating each of the two 520-nm green laser diodes. The two modulated green light beams are converted into two different OAM beams with $\ell=+1$ and $+3$ by adding different spiral phase patterns using SLMs. The generated OAM beams are coaxially combined and then split into two identical copies. Another two beams with opposite $\ell$ values of -1 and -3 can then be obtained by reflecting one of the copies three times. We note that this beam copy is relatively delayed with respect to the original one in free-space for data sequence correlation. Subsequently, the resulting four beams are spatially multiplexed and then propagated through the above-mentioned water conditions. At the receiver, each of the four OAM channels is sequentially demultiplexed using another SLM and detected using a high-sensitivity silicon avalanche photodiode (APD) with 1-GHz bandwidth. The detected signal is amplified, filtered and sent to a 1-Gbit/s receiver for bit-error rate (BER) measurements (see Supplementary Section 2 for implementation details).

Figure 3(a) depicts the eye diagrams of the 1-Gbit/s OOK signal for OAM channel $\ell=+3$ under various conditions when the other channels ($\ell=-3, -1,$ and $+1$) are turned off or on. The inter-channel crosstalk effects can be clearly observed in Figs. 3(a4-a6). In the presence of a thermal gradient, the eye diagram of channel $\ell =+3$ is time-varying due to fluctuations in the received power and crosstalk, and is not shown here. Figure 3(b) shows measured BERs as a function of received power for all four channels with and without water. The BER curve for the back-to-back (B2B) 1-Gbit/s signal is also provided as a benchmark. We observe that tap water introduces power penalties of less than 2.9 dB at the forward error correction (FEC) limit of $3.8\times10^{-3}$ for all channels. Figure 3(c) presents BER curves for OAM channels $\ell =+1$ and $+3$ under various conditions. Power penalties are measured to be 2.2, 2.3, and 2.7 dB in the cases of tap water, Maalox-induced scattering and current, respectively. Due to the effects of thermal gradient-induced turbulence, the BERs are all above the FEC limit, exhibiting a severe error-floor phenomenon, and power penalties are above 12 dB for all channels.

**40-Gbit/s OAM link using PPLN-based frequency doubling:** Due to water absorption, underwater optical communication links generally use blue-green light. However, data modulation technologies in this spectral region tend to have much lower bandwidths (e.g., around 1 GHz) than are available for IR light (e.g., beyond 10 GHz)[9, 36]. An important goal would be the achievement of higher data rates for each underwater OAM channel. Therefore, modulating data in the IR region at a much higher speed and then wavelength converting it into the blue-green region for subsequent OAM generation and underwater transmission might enable significantly higher system capacities. Specifically, whereas we previously described data rates on each OAM beam of 1 Gbit/s, we show here the ability to transmit 10 Gbit/s on each beam using frequency doubling (see Supplementary Section 3 for implementation details).

A 10-Gbit/s OOK signal at 1064 nm is generated using a lithium niobate modulator and then amplified with a high power ytterbium-doped fibre amplifier (YDFA). The 1064-nm light after amplification is sent to a frequency-doubling module that consists of a PPLN crystal and a temperature stabilized crystal oven for frequency doubling. As a result, a 532-nm green light carrying a 10-Gbit/s data stream is generated, where its power depends upon both the oven temperature and the input pump power. The generated green light acts as a light source, being converted into OAM beams using specially-designed efficient dielectric metasurface phase masks[37, 39]. Each phase mask is composed of a large number of square cross-section nano-posts that locally modify light's phase with subwavelength spatial resolution. Phase masks of $\ell = \pm 1$ and $\ell = \pm 3$ each having a blazed grating 'fork' phase pattern (i.e., combination of the spiral phase structure of the desired OAM mode and a linear phase ramp[18]) are designed, fabricated, and characterized (see Supplementary Section 4).

Employing a setup similar to the one described in the previous section, the four OAM beams with $\ell = \pm 1$ or $\pm 3$ are spatially combined and propagate through the underwater channel. At the receiver, each of the four OAM data channels is sequentially demultiplexed using a metasurface phase mask with an inverse spiral phase pattern. The beam of the desired channel is spatially filtered after demultiplexing, detected using a high-bandwidth APD (3-dB cut-off frequency of 9 GHz) and sent to a 10-Gbit/s receiver for BER measurements.

Figure 4(a) depicts the eye diagrams of the 10-Gbit/s OOK signal for OAM channel $\ell = +3$ when the other channels are turned off and on. The total crosstalk from all the other channels are -11.2, -10.7, -11.0 dB for the cases of tap water, current, and Maalox scattering, respectively. Because of this, the quality of the eye diagrams degrades when other channels are turned on. Figure 4(b) shows measured BER curves for OAM channels $\ell = +1$ and $+3$ in the cases of tap water and current with and without crosstalk from the other channels. The B2B BER curve of the 10-Gbit/s signal is also provided. The power penalties are observed to be less than 2.2 dB for all cases when all channels are on.

**Mitigation of thermal gradient-induced crosstalk using multi-channel equalisation**

Previous sections found various OAM beam degradations and consequent data-channel crosstalk based on underwater effects. In this section, we address the data degradation problem and show the mitigation of inter-channel crosstalk due to thermal gradient-induced turbulence. We employ a constant modulus algorithm (CMA)-based multi-channel equalisation in the receiver DSP to reduce channel crosstalk effects and thus recover the transmitted data streams[40-42]. This approach has been previously employed in few-mode and multi-mode fibre-based mode division multiplexed systems to mitigate the mode coupling effects among multiple spatial modes[43, 44]. In general, it is required that all the transmitted channels are simultaneously detected to enable multi-channel equalisation processing. Due to receiver hardware limitations, we only show crosstalk mitigation between two OAM channels.

With a similar system approach, two OAM beams with $\ell$ =+1 and +3 are generated using metasurface phase masks, spatially combined using a beam splitter and transmitted through water with a thermal gradient of 0.2 $^o$C. Each OAM beam carries a 10-Gbit/s OOK signal generated by doubling the frequency of a modulated 1064-nm signal using a PPLN nonlinear crystal. After demultiplexing and detection, the two OAM channels are simultaneously received, converted into Gaussian-like beams and detected by two 9-GHz bandwidth APDs. The two signals are then amplified, sampled by a real-time scope and recorded for offline DSP. A 2×2 CMA equalisation algorithm is implemented in the DSP to recover two data OAM channels with $\ell$ =+1 and $\ell$ =+3. For a 2×2 CMA equalisation, the equaliser includes four adaptive finite-impulse-response (FIR) filters each with a tap number of 11, the coefficients of which can be adaptively updated until convergence based on the CMA algorithm (see Supplementary Section 5). The obtained FIR filter coefficients are used to equalise the crosstalk between the two OAM channels.

Figure 5(a) depicts the received power and crosstalk of OAM channels $\ell =+1$ and $+3$ measured every 2 seconds under the effects of thermal gradient-induced turbulence. The received power and crosstalk fluctuate by up to 4.5 and 12.5 dB, respectively. The corresponding BERs for the two OAM channels during the same time period are shown in Fig. 5(b). Without CMA equalisation, the measured BERs fluctuate significantly between $1.7 \times 10^{-2}$ and $7.4 \times 10^{-6}$, and dramatically decrease, reaching below the FEC limit of $3.8 \times 10^{-3}$ after 2×2 CMA equalisation. We note that only a length of 2,000,000 symbols is recorded for each data sequence due to the limited memory of the real-time scope, and therefore the minimum BERs that can be measured are around $5 \times 10^{-7}$. To further illustrate the improvement, Fig. 5(c) shows the measured BERs averaged over 1 minute as a function of received power for channels $\ell =+1$ and $+3$. Due to inter-channel crosstalk, the measured BER curves without 2×2 equalization also have BER error floors. The power penalties at the FEC limit, compared to the B2B case, fall below 2.0 dB for the two channels after equalisation.

**Discussion**

The experiments described in this paper explore the potential of using OAM-based SDM to increase the transmission capacity of underwater optical communications, and several issues lend themselves to further exploration.

In general, the use of OAM multiplexing would likely require a more precise alignment between the transmitter and receiver compared to a single-channel underwater optical link. This is due to the fact that orthogonality among OAM channels relies on a common optical axis, and any misalignment may result in inter-channel crosstalk[45]. Given the beam wander that is introduced by thermal gradients, the above problem is exacerbated and will likely require an accurate pointing and tracking system.

Additionally, given that small thermal gradients can produce system degradation, we assume that this problem could become more severe for longer links for which different types of water may exist. Meanwhile, this problem may depend on the transmission direction, such that a vertical link may experience a different thermal gradient than a horizontal link. Furthermore, a channel equalisation algorithm was utilised to help mitigate the thermal gradient-induced crosstalk. However, it might be necessary under harsher and wide-ranging underwater conditions to explore the use of multiple mitigation techniques, including adaptive optics compensation and advanced channel coding[46-48]. We emphasise that other effects, such as spatial dispersion and object obstructions, are not considered yet might cause beam spreading and link outage[49, 50].

We investigated the effects of underwater propagation on OAM-multiplexed data transmission and the mitigation of inter-channel crosstalk over a short link of metre-length scale. However, we believe our results could potentially be expanded to longer distances and scaled to a larger number of OAM channels through careful system design [45] and the use of proper mitigation approaches for channel degradation effects. We envision that the underwater transmission capacity of 40 Gbit/s achieved in this paper could be further extended into sub-Tbit/s by including other techniques, such as advanced modulation formats (e.g., quadrature-amplitude-modulation and OFDM) and wavelength division multiplexing.

**Methods Summary**

**Generation and detection of data-carrying green OAM beams.** Two different data-modulation approaches are employed to generate high-speed green light signals:

*1-Gbit/s signal generation at green using internal modulation*—By directly modulating the driving current of a 520 nm laser diode, a 1-Gbit/s signal at 520 nm is produced. Due to the bandwidth limitation of the internal modulation of the laser diode, the maximal data rate of the green beam is 1 Gbit/s. The generated signal is then launched onto a programmable SLM with a specific helical phase pattern to create an OAM beam with either $\ell$ =+1 or +3. Multiple generated OAM beams are then multiplexed using a beam splitter-based combiner and the resulting beams propagate through the underwater channel. The received signal after demultiplexing is detected using a high sensitivity Si APD with a 3-dB bandwidth of 1-GHz.

*10-Gbit/s signal generation at green using PPLN-based frequency doubling*—Generally, the modulation bandwidth of both internal and external modulations for green light is limited to GHz [4, 7]. To overcome this, the frequency doubling of a data-carrying 1064-nm signal is thus used to produce a high-speed green light signal. Specifically, we perform high-speed data modulation using a 1064-nm lithium niobate modulator and use a PPLN-based frequency-doubling module to convert the carrier wavelength from 1064 nm to 532 nm. Consequently, a green light signal at 532 nm is generated, which is then split into multiple copies and converted into OAM beams using transmissive metasurface phase masks. At the receiver, a Si APD with a 3-dB bandwidth of 9-GHz but a lower sensitivity than the detector used for the 520-nm signal detection is employed for signal detection.

**Crosstalk mitigation using multi-channel equalisation.** The multiplexed OAM beams may be distorted due to underwater propagation, causing the power spreading of each transmitted OAM mode onto neighbouring modes. Consequently, each OAM channel experiences interferences from the other channels, resulting in a non-diagonal channel matrix. Theoretically, to recover the data streams, the received signals of all OAM channels could then be multiplied with the inverse channel matrix. In our experiment, we use a 2×2 CMA adaptive channel equalisation in the receiver to reduce the effects of interferences and recover the two data channels. In general, the dimension of the equalisation processing is determined by the total number of OAM channels. The CMA-based equalisation utilises an FIR filter-based linear equaliser for each channel. The FIR-CMA equaliser contains four FIR filters, the coefficients of which can be adaptively updated until convergence based on the CMA. The obtained FIR filter coefficients are used to equalise the crosstalk between two OAM channels.

**References**

1. Akyildiz, I. F. Pompili, D. & Melodia, T. Underwater acoustic sensor networks: research challenges. *Ad Hoc Netw.* **3**, 257–279 (2005).

2. Stojanovic, M. Recent advances in high-speed underwater acoustic communication. *IEEE J. Oceanic Eng.* **21**, 125–136 (1996).

3. Arnon, S. Underwater optical wireless communication network. *Opt. Eng.* **49**, 015001 (2010).

4. Munafo, A. Simetti, E. Turetta, A. Caiti, A. & Casalino, G. Autonomous underwater vehicle teams for adaptive ocean sampling: a data-driven approach. *Ocean Dyn.* **61**, 1981–1994 (2011).

5. Lacovara, P. High-bandwidth underwater communications. *Mar. Technol. Soc. J.* **42**, 93–102 (2008).

6. Simpson, J. A. Hughes, B. L. & Muth, J. F. Smart transmitters and receivers for underwater free-space optical communication. *IEEE J. on Sel. Areas in Commun.* **30**,

964–974 (2012).

7. Baiden, G. Bissiri, Y. & Masoti, A. Paving the way for a future underwater omni-directional wireless optical communication systems. *Ocean Eng.* **36**, 633–640 (2009).

8. Hanson, F. & Radic, S. High bandwidth underwater optical communication. *Appl. Opt.* **47**, 277–283 (2008).

9. Nakamura, K. Mizukoshi, I. & Hanawa, M. Optical wireless transmission of 405 nm, 1.45 Gbit/s optical IM/DD-OFDM signals through a 4.8 m underwater channel. *Opt. Express* **23**, 1558–1566 (2015).

10. Oubei, H. M. *et al*. 2.3 Gbit/s underwater wireless optical communications using directly modulated 520 nm laser diode. *Opt. Express* **23**, 20743–20748 (2015).

11. Oubei, H. M. *et al*. 4.8 Gbit/s 16-QAM-OFDM transmission based on compact 450-nm laser for underwater wireless optical communication. *Opt. Express* **23**, 23302-23309 (2015).

12. Johnson, L. Green, R. & Leeson, M. S. Underwater optical wireless communications: depth dependent variations in attenuation. *Appl. Opt.* **52**, 7867–7873 (2013).

13. Johnson, L. Green, R. & Leeson, M. S. Underwater optical wireless communications: depth-dependent beam refraction. *Appl. Opt.* **53**, 7273-7277 (2014).

14. Cox W. & Muth, J. Simulating channel losses in an underwater optical communication system. *J. Opt. Soc. Am. A* **31**, 920–934 (2014).

15. Jaruwatanadilok, S. Underwater wireless optical communication channel modeling and performance evaluation using vector radiative transfer theory. *IEEE J. Sel. Areas Commun.* **26**, 1620–1627 (2008).

16. Gibson, G. *et al*. Free-space information transfer using light beams carrying orbital

angular momentum. *Opt. Express* **12,** 5448–5456 (2004).

17. Allen, L., Beijersbergen, M. W., Spreeuw, R. J. C. & Woerdman, J. P. Orbital angular-momentum of light and the transformation of Laguerre-Gaussian laser modes. *Phys. Rev. A* **45,** 8185–8189 (1992).

18. Yao, A. M. & Padgett, M. J. Orbital angular momentum: origins, behavior and applications. *Adv. Opt. Photon.* **3,** 161–204 (2011).

19. Martelli, P. Gatto, A. Boffi, P. & Martinelli, M. Free-space optical transmission with orbital angular momentum division multiplexing. *Electron. Lett.* **47**, 972–973 (2011).

20. Wang, J. *et al.* Terabit free-space data transmission employing orbital angular momentum multiplexing. *Nat. Photonics* **6,** 488–496 (2012).

21. Huang H. *et al.* 100 Tbit/s free-space data link enabled by three-dimensional multiplexing of orbital angular momentum, polarization, and wavelength. *Opt. Lett.* **39**, 197-200 (2014).

22. Ren, Y. *et al.* Experimental characterization of a 400 Gbit/s orbital angular momentum multiplexed free-space optical link over 120 m. *Opt. Lett.* **41**, 622-625 (2016).

23. Zhao, Y. *et al.* Experimental Demonstration of 260-meter Security Free-Space Optical Data Transmission Using 16-QAM Carrying Orbital Angular Momentum (OAM) Beams Multiplexing. *Proc. of Optical Fiber Communication Conference*, paper Th1H.3 (2016).

24. Tyler, G. A. & Boyd, R. W. Influence of atmospheric turbulence on the propagation of quantum states of light carrying orbital angular momentum. *Opt. Lett.* **34**, 142-144 (2009).

25. Anguita, J. A. Neifeld, M. A. & Vasic, B. V. Turbulence-induced channel crosstalk in an orbital angular momentum-multiplexed free-space optical link. *Appl. Opt.* **47**, 2414-2429 ( 2008).

26. Ren, Y. *et al.* Atmospheric turbulence effects on the performance of a free space optical link employing orbital angular momentum multiplexing. *Opt. Lett.* **38**, 4062-4065 (2013).

27. Krenn, M. *et al.* Communication with spatially modulated light through turbulent air across Vienna. *New J. Phys.* **16**, 1-10 (2014).

28. Baghdadya, J. *et. al.* Spatial multiplexing for blue lasers for undersea communications. *Proc. of SPIE* **9459**, 1-7 (2015).

29. Morgan, K. S. Johnson, E.G. & Cochenour, B.M. Attenuation of beams with orbital angular momentum for underwater communication systems. IEEE MTS OCEANS, 1–3 (IEEE 2015).

30. Ata, Y. & Baykal, Y. Structure functions for optical wave propagation in underwater medium. *Waves in Random and Complex Media* **24**, 164–173 (2014).

31. Cochenour, B. Mullen, L. & Muth, J. Effect of scattering albedo on attenuation and polarization of light underwater. *Opt. Lett.* **35**, 2088–2090 (2010).

32. Cochenour, B. Mullen, L. & Laux, A. E. Characterization of the beam-spread function for underwater wireless optical communications links. *J. Oceanic Eng.* **33**, 513–521 (2008).

33. Hanson F. & Lasher, M. Effects of underwater turbulence on laser beam propagation and coupling into single-mode optical fiber. *Appl. Opt.* **49**, 3224–3230 (2010).

34. Yi, X. Li, Z. & Liu, Z. Underwater optical communication performance for laser beam propagation through weak oceanic turbulence. *Appl. Opt.* **54**, 1273-1278 (2015).

35. Ren, Y. *et. al.* 4 Gbit/s Underwater optical transmission using OAM multiplexing and directly modulated green laser. *OSA Conference on Lasers and Electro-Optics (CLEO)*, paper SW1F.4, 1-2 (2016).

36. Mullen, L. Cochenour, B. Laux, A. & Alley D. Optical modulation techniques for underwater detection, ranging and imaging. *Proc. of SPIE*, 8030, 1-9 (2011).

37. Arbabi, A. Horie, Y. Bagheri, M, & Faraon, A. Dielectric metasurfaces for complete control of phase and polarization with subwavelength spatial resolution and high transmission. *Nat. Nanotech.* **10**, 937–943 (2015).

38. Li, L. *et. al.* CMA equalization for a 2 Gb/s orbital angular momentum multiplexed optical underwater link through thermally induced refractive index inhomogeneity. *OSA Conference on Lasers and Electro-Optics (CLEO)*, paper SW1F.2, 1-2 (2016).

39. Arbabi, A. Horie, Y. Ball, A. J. Bagheri, M, & Faraon, A. Subwavelength-thick lenses with high numerical apertures and large efficiency based on high-contrast transmitarrays. *Nat. Comm.* **6**, 1–6 (2015).

40. Ready, M. & Gooch, R. Blind equalization on radius directed adaptation. *Proc. of International Conference on Acoustics, Speech, and Signal Processing (ICASSP-90)*. **3**, 1699 (IEEE 1990).

41. Treichler, J. R. & Agee, B. G. A new approach to multipath correction of constant modulus signals. *IEEE Trans. on Acoustics, Speech, and Signal Processing* **31**, 459–472 (2004).

42. Huang, H. *et. al.* Crosstalk mitigation in a free-space orbital angular momentum multiplexed communication link using 4×4 MIMO equalization. *Opt. Lett.* **39**, 4360-4363 (2014).

43. Dar, R. Feder, M. & Shtaif, M. The underaddressed optical multiple-input, multiple-output channel: capacity and outage. *Opt. Lett.* **37**, 3150-3152 (2012).

44. Richardson, D. J. Fini, J. M. & Nelson, L. E. Space-division multiplexing in optical fibres.

*Nat. Photonics* **7**, 354–362 (2013).

45. Xie, G. *et. al.* Performance metrics and design considerations for a free-space optical orbital-angular-Momentum multiplexed communication link. *Optica* **2**, 357-365 (2015).

46. Djordjevic I. B. & Arabaci, M. LDPC-coded orbital angular momentum (OAM) modulation for free-space optical communication. *Opt. Express* **18**, 24722-24728 (2010).

47. Ren, Y. *et al.* Adaptive-optics-based simultaneous pre- and post-turbulence compensation of multiple orbital-angular-momentum beams in a bidirectional free-space optical link. *Optica* 1, 376-382 (2014).

48. Rodenburg, B. *et al.* Simulating thick atmospheric turbulence in the lab with application to orbital angular momentum communication. *New J. Phys.* **16**, 1-12 (2014).

49. Ahmed, N. *et. al.* Mode-division-multiplexing of multiple Bessel-Gaussian beams carrying orbital-angular-momentum for obstruction-tolerant free-space optical and millimetre-wave communication links. *Scientific Reports* **6**, 1-7 (2016).

50. Cochenour, B. Mullen, L. & Laux, A. Spatial and temporal dispersion in high bandwidth underwater laser communication links. *Proc. IEEE Military Communications Conf.* 1–7 (IEEE 2008).

**Acknowledgements**

We acknowledge James M. Krause, Michael J. Luddy, and Jack H. Winters for valuable help and fruitful discussions. S.M.K. is supported by 'Light-Material Interactions in Energy Conversion' Energy Frontier Research Center funded by the US Department of Energy, Office of Science, Office of Basic Energy Sciences. E.A. and A.A. are supported by Samsung Electronics. Device nanofabrication was performed at the Kavli Nanoscience Institute at California Institute of Technology. This work is supported by the National Science Foundation and NxGen Partners.

**Author contributions**

Y.R., L.L., and A.W. developed the concept and designed the experiments. Y.R., L.L., Z.Z., G.X., Z.W., N.A., Y.Y., C.L., and A.J.W carried out the measurements and analysed the data. L.L., Y.C., and Y.R. designed and implemented the multiple-input-multiple-output equalisation algorithm. S. M. K., E. A., A. A., and A. F. designed, fabricated, and characterized the metasurface OAM generator phase masks. S.A., M.T., A.F., and A.W. provided technical support. The project was conceived and supervised by A.W..

The authors declare no competing financial interests. Correspondence and requests for materials should be addressed to A.W. (willner@usc.edu).

**Figure 1**

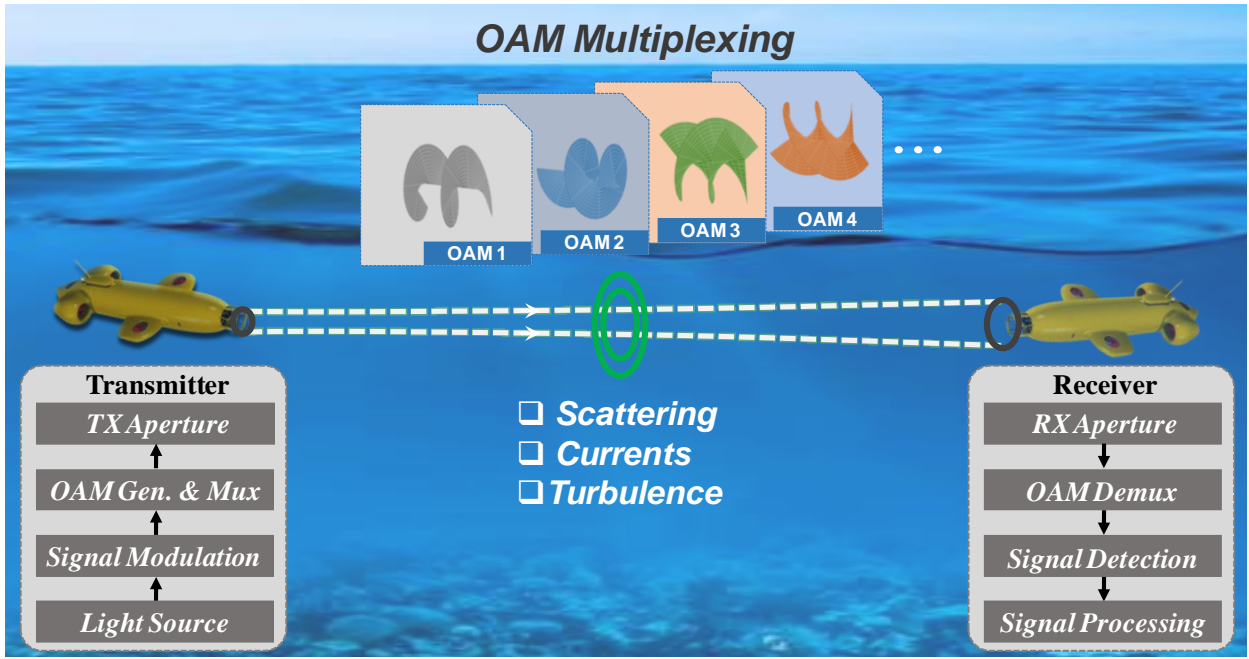

**Figure 1 | Prospective application scenario for a high-capacity underwater optical communications link with OAM-based space division multiplexing**. Key modules including light source, signal modulation, OAM generation/multiplexing, OAM demultiplexing/detection and receiver signal processing are shown.

**Figure 2**

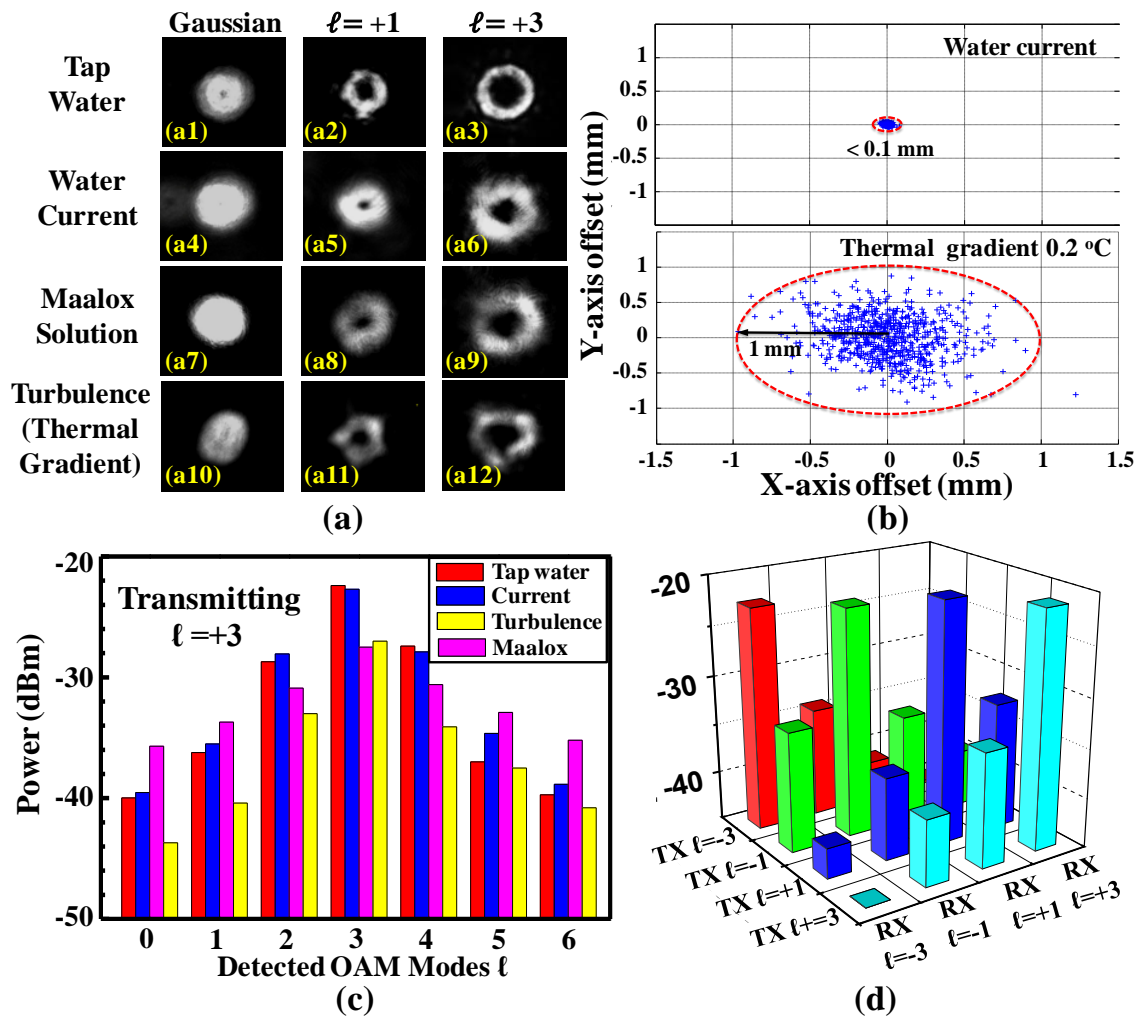

**Figure 2 | OAM beam propagation through various underwater channel conditions.** (a) Intensity profiles of OAM beams under various channel conditions: (a1-a3) with only tap water, (a4-a6) with water current, (a7-a9) with the Maalox solution, and (a10-a12) with thermal gradient-induced turbulence. (b) Statistics for beam wander at the receiver with respect to the propagation axis due to water current and thermal gradient-induced turbulence. (c) OAM power spectrum when transmitting OAM channel $\ell = +3$ under various conditions. (d) Power transfer between all OAM channels under water current.

**Figure 3**

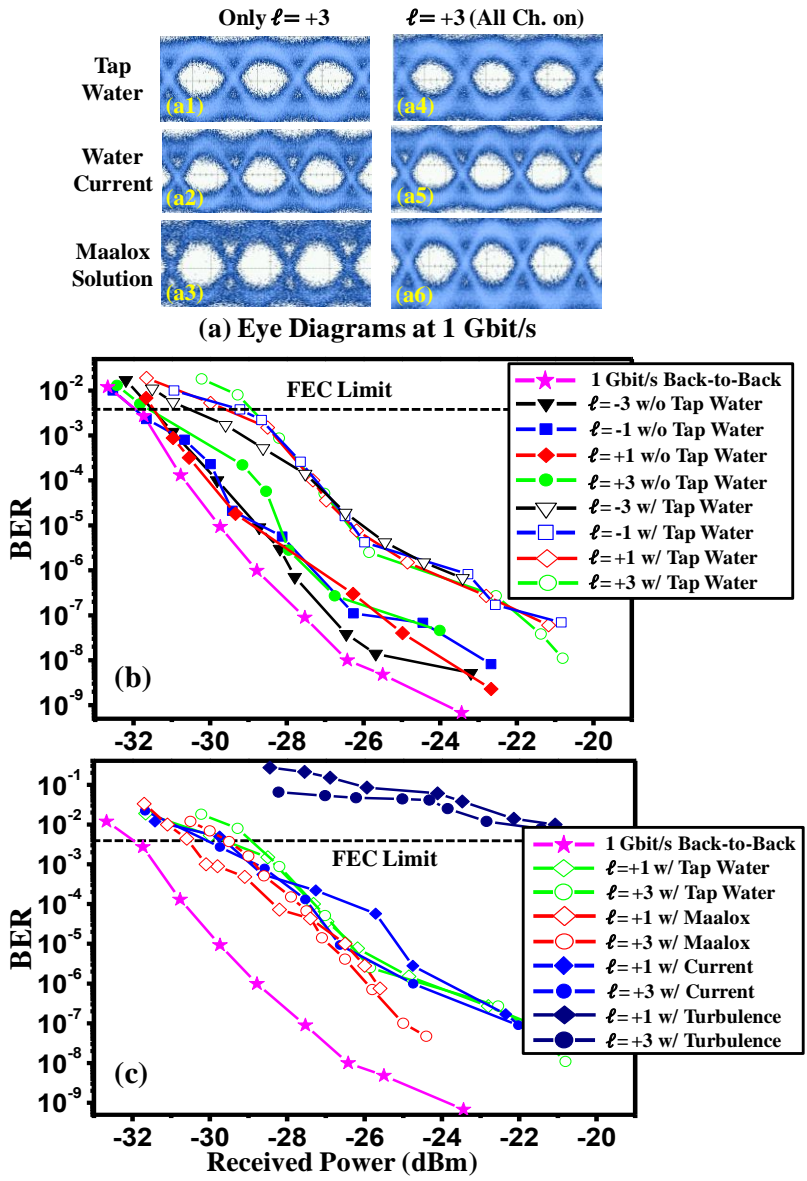

**Figure 3 | System performance measurements for the 4-Gbit/s underwater link using directly modulated laser diodes.** (a) Eye diagrams for OAM channel $\ell$ =+3, (b) BERs as a function of received power with and without tap water, and (c) BERs with Maalox-induced scattering, current and thermal gradient-induced turbulence.

**Figure 4**

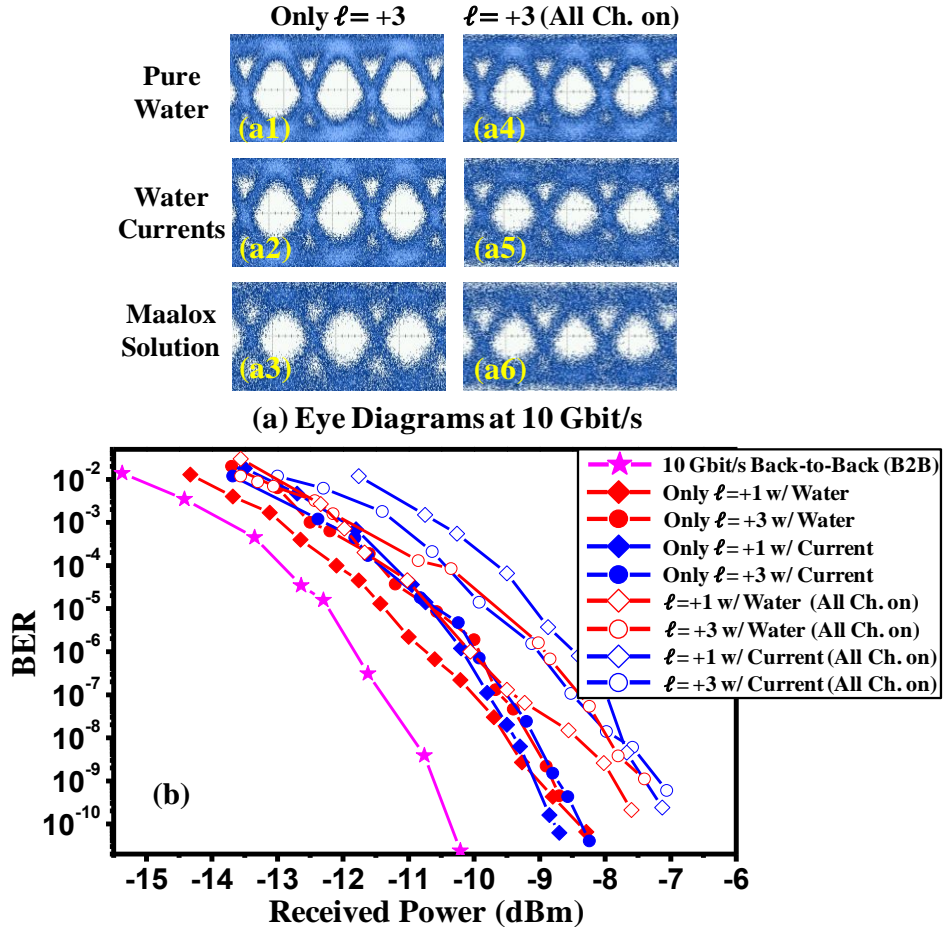

**Figure 4 | System performance measurements for the 40-Gbit/s underwater link using PPLN-based frequency doubling for signal generation.** (a) Eye diagrams for OAM channel $\ell$ =+3 at a fixed transmitted power when other channels are turned off or on, and (b) BERs as a function of received power with tap water and current.

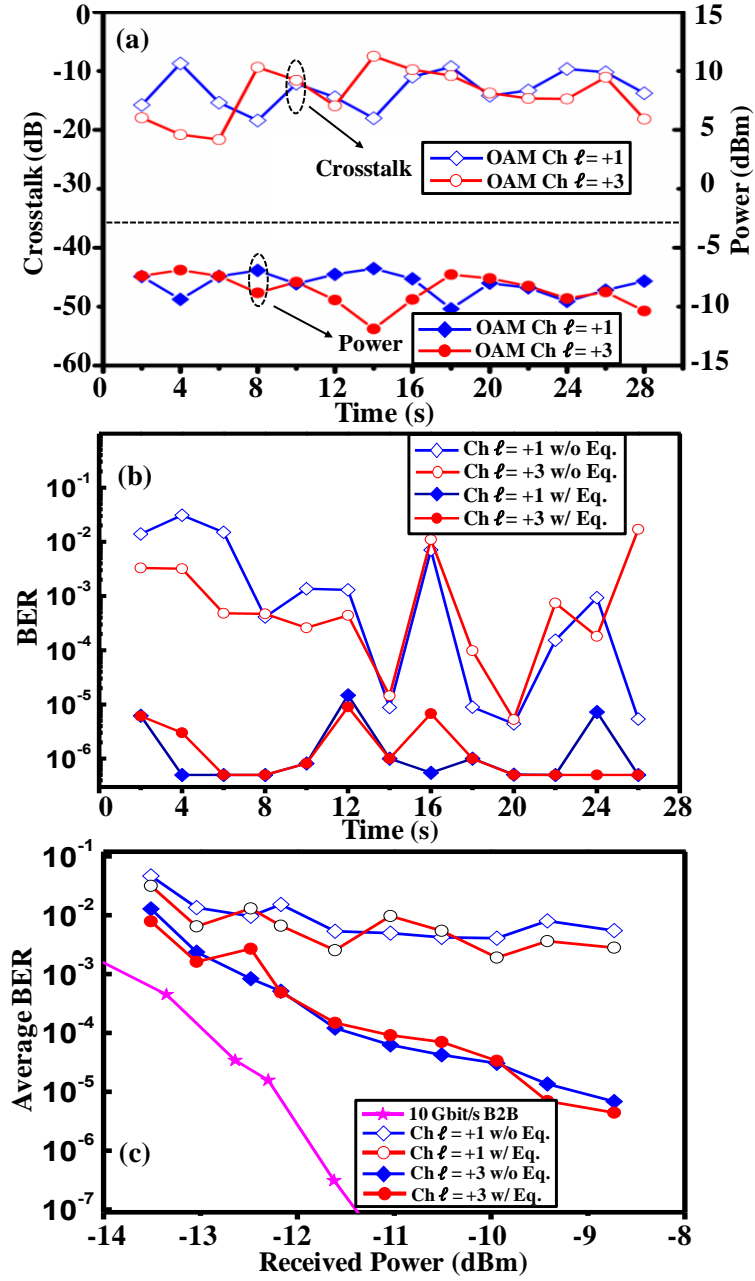

**Figure 5 | Mitigation of thermal gradient-induced crosstalk using CMA-based multi-channel equalisation.** (a) Received power and channel crosstalk of OAM $\ell$ =+1 and $\ell$ =+3 over 28 seconds and (b) instantaneous BER of OAM $\ell$ =+1 and $\ell$ =+3 over 28 seconds with and without CMA equalisation under thermal gradient-induced turbulence when both channels are

transmitted. (c) Measured BER curves of OAM channel $\ell=+1$ and $\ell=+3$ with and without CMA equalisation. Ch.: channel. Eq.: equalization.

# *Orbital Angular Momentum-based Spatial Division Multiplexing for High-capacity Underwater Optical Communications*

Yongxiong Ren[1]*, Long Li[1]*, Zhe Wang[1], Seyedeh Mahsa Kamali[2], Ehsan Arbabi[2], Amir Arbabi[2], Zhe Zhao[1], Guodong Xie[1], Yinwen Cao[1], Nisar Ahmed[1], Yan Yan[1], Cong Liu[1], Asher J. Willner[1], Solyman Ashrafi[3], Moshe Tur[4], Andrei Faraon[2], and Alan E. Willner[1]

[1]Department of Electrical Engineering, University of Southern California, Los Angeles, CA 90089, USA.

[2]T. J. Watson Laboratory of Applied Physics, California Institute of Technology, Pasadena, CA 91125, USA

[3]NxGen Partners, Dallas, TX75219, USA

[4]School of Electrical Engineering, Tel Aviv University, Ramat Aviv 69978, Israel.

*These authors contributed equally to this work.

## 1. The underwater propagation of OAM beams: Measurements under a different scattering level and thermal gradient

We show below measurements of OAM beam propagation under a different scattering level and thermal gradient. Supplementary Figs. S1(a1–a2) show BER fluctuations over 36 seconds at a fixed transmitted power of -26 dBm when different amounts of 0.5% diluted Maalox solution are added to the water. These measurements are performed every two seconds and repeated 18 times before the Maalox particles are evenly distributed in the water (i.e., before circulation over 1 minute). We see that the case of 0.5-millilitre of Maalox solution has a larger BER fluctuation range due to a more uneven suspension of Maalox particles along the link path. After uniform scattering suspension is obtained, the 0.5- and 1.5-millilitre Maalox solutions cause power losses of 2.2 dB and 4.5 dB to the link, respectively. Figure S1(c) presents the statistics of beam displacement with respect to the propagation axis after propagation through thermal gradient-induced turbulence, in which the room temperature water and heated water have a temperature difference of 0.3$^o$C. We see that the received beams' maximal displacement is estimated to be 1.4 mm, which is larger than that shown in Fig. 2(b) in the main manuscript. The captured intensity profiles for OAM beam $\ell=+3$ under three different turbulence realisations are also shown to illustrate the time-varying distortions caused by dynamic turbulence. Figure S1(d) shows the BERs and received power for the OAM $\ell=+3$ channel under various turbulence realisations. A wider range of power fluctuation is observed than that shown in Fig. 5(a) in the main manuscript.

## 2. Implementation details of the 4-Gbit/s OAM multiplexed underwater optical link using directly modulated green laser diodes

In this section, we describe the experimental implementation of the 4-Gbit/s four OAM multiplexed underwater link, in which laser diodes at 520 nm are directly modulated and act as light signal sources. Supplementary Fig. S2 presents the link schematic. Two 1-Gbit/s OOK signal beams at 520 nm are generated by directly modulating each of the two 520-nm green laser diodes with a binary sequence. The two modulated green light beams are launched onto two liquid crystal-based spatial light modulators (SLMs) to create two different OAM beams with $\ell=+1$ and +3. The SLM (Santec Inc.) has a pixel resolution of 1440×1050 and an operating wavelength range of 500-1650nm. These two OAM beams $\ell=+1$ and +3 are coaxially combined and then split into two identical copies, one of which is reflected three times using mirrors arranged to introduce a ~50 ns delay for data sequence decorrelation between the two copies. Another two OAM beams with opposite $\ell$ values of -1 and -3 are obtained due to the odd number of reflections; these new beams are then combined with the original OAM beams ℓ = +1 and +3 by a beam splitter. The resulting four multiplexed OAM beams propagate through the underwater channel.

**(a) Multiplexed OAM beams at the transmitter**

| dBm | ℓ= -3 | ℓ= -1 | ℓ= +1 | ℓ= +3 |
|---|---|---|---|---|
| ℓ= -3 | -23.72 | -34.12 | -40.82 | -44.53 |
| ℓ= -1 | -35.08 | -21.42 | -36.52 | -39.46 |
| ℓ= +1 | -40.60 | -32.89 | -21.78 | -33.06 |
| ℓ= +3 | -40.13 | -38.43 | -32.58 | -20.91 |

**(b) After propagating through tap water**

| dBm | ℓ= -3 | ℓ= -1 | ℓ= +1 | ℓ= +3 |
|---|---|---|---|---|
| ℓ= -3 | -23.39 | -34.01 | -39.90 | -42.87 |
| ℓ= -1 | -36.80 | -21.89 | -32.45 | -39.23 |
| ℓ= +1 | -39.20 | -34.12 | -22.67 | -34.50 |
| ℓ= +3 | -42.80 | -39.26 | -34.89 | -22.35 |

**Table 1.** Power transfer (dBm) between the four multiplexed OAM channels (a) at the transmitter and (b) after propagating through tap water.

We characterise the power leakage and crosstalk between all OAM channels. The power leakage is measured as follows: First we transmit a 520-nm signal over the OAM channel ℓ=-3 while all the other channels (OAM beams ℓ=-1, +1, and +3) are off. Then we record the received power for ℓ=-3, as well as the power leaked into other OAM modes (ℓ =±1 and +3). The above measurements are repeated for all transmitted OAM channels until a full 4×4 power transfer matrix is obtained. Table 1 shows the power transfer matrices between all four channels at the transmitter and after propagating the tap water channel. The crosstalk of a specific channel can be calculated from the 4×4 power transfer matrix by adding the received power from all other channels divided by the received power of the channel under consideration. We see that the crosstalk values for all four multiplexed channels are below -10 dB in both cases, and that the degradation introduced by the tap water is negligible.

Three different underwater channel conditions are emulated in a 1.2-metre-long rectangular tank filled with tap water, as described in the main manuscript. The water current and scattering are created by distributed circulation pumps and the Maalox solution, respectively. The thermal gradient-induced turbulence is produced by mixing the room temperature and heated water. After propagating through the various water conditions, the four multiplexed OAM channels are sequentially demultiplexed and detected at the receiver. To recover the data channel carried by the OAM beam $+\ell$, the SLM used for demultiplexing is loaded with a inverse spiral phase pattern of $-\ell$. As a result, only the OAM beam with $+\ell$ is converted into a Gaussian-like beam ($\ell = 0$) while all other beams maintain ring-shaped profiles, which could be efficiently filtered out by a spatial filter (simply a pin hole). The Gaussian-like beam is subsequently focused onto a high-sensitivity silicon avalanche photodiode detector (APD) with a 3-dB bandwidth of 1 GHz. The APD has a spectral responsivity of ~15 $A/W$ at 520 nm and a low noise equivalent power (NEP) of 0.4 $pW/\sqrt{Hz}$. After detection, the signal is amplified, filtered and sent to a 1-Gbit/s receiver for bit-error rate (BER) measurements.

## 3. Implementation details of the 40-Gbit/s OAM multiplexed underwater optical link using PPLN-based frequency doubling

We present below the implementation details of the 40-Gbit/s four OAM multiplexed underwater link. In this experiment, each 532-nm green OAM beam carries a 10-Gbit/s on-off-keyed (OOK) signal, which is generated using frequency doubling based on a periodically poled lithium niobate (PPLN) nonlinear crystal. The experimental setup is depicted in Supplementary Fig. S3.

A 1064-nm single-mode laser with a linewidth of less than 100 MHz is sent to a lithium niobate modulator to produce a 10-Gbit/s OOK signal. The transmitted RF signal is a pseudorandom binary sequence with a length of $2^{15}$-1. The 10-Gbit/s signal at 1064 nm is amplified by a ytterbium-doped fibre amplifier (YDFA) and fed into a collimator to convert the single-mode fibre output to a collimated Gaussian beam with a diameter of 2.6 mm. This beam is then focused into the centre of a PPLN crystal for frequency doubling. A half-wave plate is inserted after the collimator to align the polarisation of incident light with the polarisation orientation of the PPLN crystal to maximise the conversion efficiency. The PPLN crystal is z-cut to 20 mm × 1 mm × 1 mm in dimension. Both its input and output facets are dual-coated with reflectivity of <1% at 532 nm and 1064 nm to reduce the Fresnel reflection loss. The crystal oven together with a temperature controller offers a stability of ±0.01°C. The power of the green beam generated from frequency doubling depends upon both the temperature and the input pump power. For a 34 dBm 1064-nm input beam, the power of the output green beam at 532 nm is around 22 dBm with the oven temperature being 79.5 °C. The beam after the PPLN crystal is collimated and passes through a dichroic mirror to reflect the unconverted 1064-nm beam. A bandpass filter with a central wavelength of 532 nm is followed to further separate the green light from the remaining 1064-nm beam. As a result, a 532-nm green beam carrying a 10-Gbit/s OOK signal is produced. We note that this frequency-doubling process is transparent to the intensity-based modulation format.

The resulting signal beam at 532 nm is split into two copies, which then pass through two transmissive metasurface phase masks to generate two OAM beams with $\ell$=+1 and +3, respectively. The metasurface phase masks have high efficiency (power loss of ~3dB at 532 nm) and the generated OAM modes are of high quality (see Supplementary Note 4 for more details on the design). These two OAM beams are spatially combined using a beam splitter. Using a similar approach as in the previous section, another two OAM beams with opposite ℓ values (ℓ = -1 and -3) can be obtained, which are then multiplexed with the original OAM beams ℓ = +1 and +3. The sizes of the OAM beams ℓ = -3, -1, +1 and +3 are 2.1, 1.5, 1.4 and 2.0 mm, respectively. The resulting four multiplexed OAM beams (ℓ = ±1 and ±3) are sent through a 1.2-metre underwater channel. The power transfer matrices between the four channels at the transmitter and after propagating the tap water channel are shown in Table. 2. We see that the crosstalk values for all four channels are below -11 dB in both cases.

**(a) Multiplexed OAM beams at the transmitter**

| dBm | ℓ= -3 | ℓ= -1 | ℓ= +1 | ℓ= +3 |
|---|---|---|---|---|
| ℓ= -3 | -17.3 | -29.8 | -31.2 | -39.6 |
| ℓ= -1 | -31.5 | -17.4 | -28.3 | -38.74 |
| ℓ= +1 | -38.7 | -30.9 | -17.5 | -28.3 |
| ℓ= +3 | -17.7 | -32.3 | -29.2 | -17.7 |

**(b) After propagating through tap water**

| dBm | ℓ= -3 | ℓ= -1 | ℓ= +1 | ℓ= +3 |
|---|---|---|---|---|
| ℓ= -3 | -19.99 | -31.8 | -33.65 | -40.18 |
| ℓ= -1 | -32.07 | -20.1 | -31.66 | -37.42 |
| ℓ= +1 | -35.49 | -30.1 | -19.89 | -33.12 |
| ℓ= +3 | -40.18 | -35.1 | -30.6 | -19.83 |

**Table 2.** Power transfer (dBm) between the four multiplexed OAM channels (a) at the transmitter and (b) after propagating through the tap water channel.

After propagation through water, a phase mask with an inverse phase pattern of the desired OAM channel is used to convert the chosen OAM beam into a Gaussian like beam. The other beams maintain their ring-shaped profiles and helical phases after passing through the phase mask. The Gaussian like beam has a bright high intensity at its centre and is therefore separable from the other beams through spatial filtering (simply a pin hole). This beam is then focused onto a 9-GHz-bandwidth silicon APD, which has a spectral responsivity of ~0.2 *A/W* at 532 nm and an NEP of <45 $pW/\sqrt{Hz}$. The signal is then amplified, filtered and sent to a 10-Gbit/s OOK receiver for BER measurements. Given that four OAM beams each bearing a 10-Gbit/s data stream are transmitted, a total capacity of 40 Gbit/s is achieved.

Supplementary Figs. S4(a1–a4) show the captured intensity profiles of the generated OAM beams of ℓ = ±1 and ℓ = ±3 at the transmitter. Figures S4(b1–b2) present the measured interferograms for OAM beams ℓ = +1 and +3, in which the state number of the two OAM beams can be deduced from the number of rotating arms. Each interferogram is obtained interfering an OAM beam (either ℓ = +1 or +3) with an expanded Gaussian beam. Figs. S4(c1–c4) show the measured intensity profiles (c1-c2) and interferograms (c3-c4) of the received OAM beams ℓ = +1, and ℓ = +3 after propagating through tap water. Figs. S4(d1–d4) depict the intensity profiles of demultiplexed beams at the receiver when only the OAM channel ℓ = +3 is transmitted. We see that only when the phase mask is of inverse spiral phase of ℓ = -3, can the OAM beam with ℓ = +3 be converted into a Gaussian like beam with a high intensity at centre.

## 4. High-efficiency dielectric metasurface OAM generator for visible wavelengths

In this section, we describe the design, fabrication and characterisation of dielectric metasurface phase masks for OAM generation. Such phase masks made of square cross-section SiN$_x$ nano-posts on fused silica are designed, fabricated and characterised for the generation and detection of OAM beams at 532 nm. Polarisation-insensitive phase masks are composed of 630-nm-tall SiN$_x$ nano-posts on a square lattice with the lattice constant of 348 nm. By changing the nano-posts width in the range of 60 nm to 258 nm, the transmission phase can be changed from 0 to 2π while maintaining high transmissivity at the wavelength of 532 nm. Therefore, any arbitrary phase profile can be designed using this metasurface platform. Here, a blazed grating 'fork' phase pattern, which is a combination of the helical phase structure of the desired OAM mode and a small linear phase ramp (~2 degrees), is designed. Such a combination pattern can help separate the generated OAM beam from the residual unmodulated Gaussian beam, providing high-quality OAM generation and detection. Supplementary Fig. S5(a) shows a schematic illustration of the blazed grating 'fork' phase mask generating the OAM mode with $\ell=+3$. The metasurfaces are fabricated by depositing a 630-nm thick layer of $SiN_x$ on a fused silica substrate. The pattern is defined and transferred to the $SiN_x$ layer using e-beam lithography followed by the lift-off process and dry etching. Two different 1.5-mm diameter metasurface phase masks with $\ell=+1$ and $\ell=+3$ are designed and fabricated. Figs. S5(b1–b2) show the scanning electron micrographs of the fabricated phase masks of $\ell=+1$ and $\ell=+3$, respectively. To quantify the quality of OAM beam generation, we measure the OAM power spectrum of the generated OAM beams $\ell = +1$ and $\ell = +3$, as shown in Fig. S5(c). We see that the majority of the power resides in the desired modes and the power leakage onto the adjacent modes is -13 dB less than the desired modes.

## 5. Implementations details of the CMA-based multi-channel equalisation algorithm

In general, for an underwater optical link using *M* multiplexed OAM beams, a multiple-input multiple-output (MIMO) channel processing with a dimension of $M \times M$ would be needed to reduce the inter-channel crosstalk caused by underwater propagation (mainly thermal gradient-induced turbulence in the experiments). Mathematically speaking, the received signal vector for all *M* OAM channels $\boldsymbol{y} = (y_1, y_2, ..., y_M)^\mathrm{T}$ can be expressed as

$$y = Hx + N \qquad (1)$$

where $y_i$ ($i = 1, 2, \ldots M$) is the received signal of OAM channel $i$ and $x = (x_1, x_2, \ldots, x_M)^T$ with $x_i$ being the transmitted signal for OAM channel $i$. $H$ is the channel matrix, which can be written as

$$H = \begin{bmatrix} h_{1,1} & h_{1,2} & \cdots & h_{1,M} \\ h_{2,1} & h_{2,2} & \cdots & h_{2,M} \\ \vdots & \vdots & \ddots & \vdots \\ h_{M,1} & h_{M,2} & \cdots & h_{M,M} \end{bmatrix}_{M \times M}, \qquad (2)$$

where $h_{i,j}$ depicts the transfer function between OAM channel $i$ to OAM channel $j$. $N = (n1, n2, \ldots, nM)^T$ and $ni$ is the noise for $i$-th OAM channel. $H$ is determined by the power loss of each OAM channel and crosstalk between all $M$ OAM channels, which are directly related to the underwater channel conditions and system design (e.g. aperture sizes and link distance). To recover the OAM data streams, the received signals of all OAM channels could then be multiplied with the inverse of channel matrix $H$ theoretically.

A variety of implementation approaches for MIMO processing based on $H$ have been proposed, including joint maximum likelihood sequence estimation of the data symbols in different streams, minimum mean-square error detection combined with serial interference cancellation, zero forcing detection, and pilot-aided channel equalisation. For our experiment using two multiplexed OAM modes, a 2×2 CMA-based equalisation algorithm is implemented to equalise crosstalk between the two OAM channels, thereby allowing data recovery. After demultiplexing, each of the two received OAM channels is converted into a Gaussian-like beam and detected by a 1-GHz bandwidth APD. The two signals are amplified, sampled by a real-time scope and recorded for offline DSP. Given that we use the OOK modulation format for the signal generation, which does not have a constant modulus in the signal constellation, the DC components of the two received channels are subtracted to apply the CMA algorithm. The CMA-based equalisation algorithm utilises a linear equaliser for each channel. For a 2×2 equalisation, the equaliser includes four adaptive finite-impulse-response (FIR) filters, each with a tap number of $K$. Specifically, the output of the equaliser corresponding to each channel can be expressed as:

$$y_j = \sum_{i=1}^{2} \boldsymbol{w_{ij}} * \boldsymbol{x_i}, j = 1, 2 \qquad (3)$$

where $\boldsymbol{w_{ij}}$ ($i$ =1,2) is the coefficient vector of the FIR filter with a vector length of $K$ (tap number). $\boldsymbol{w_{ij}} * \boldsymbol{x_i}$ represents the inner product operation between two vectors and $y_j$ is the output of the FIR filter. All the FIR coefficients are initialised as zero with only the centre weight being 1 and then updated until the coefficients converge based on CMA:

$$\boldsymbol{w_{ij}}(k+1) = \boldsymbol{w_{ij}}(k) + u \cdot e_i \cdot y_i \cdot \boldsymbol{x_i}^* \qquad (4)$$

where $u$ is the step size, $e_i = P_{ref} - |y_i|^2$ is the error signal of the adaptive estimation and $P_{ref}$ is the normalised reference power. The main idea of CMA-based MIMO equalisation is to update filter weights so that each channel output can have a clear amplitude. The tap number $K$ in each FIR filter is set to be 11, which is sufficient to cover the differential time delays among each data sequence and mitigate temporal ISI effects. The obtained FIR filter coefficients are used to equalise the crosstalk among two OAM channels as in Equation (3). After equalisation, the bit-error rates (BERs) are evaluated for both channels.

**Supplementary Figures**

## Supplementary Figure S1

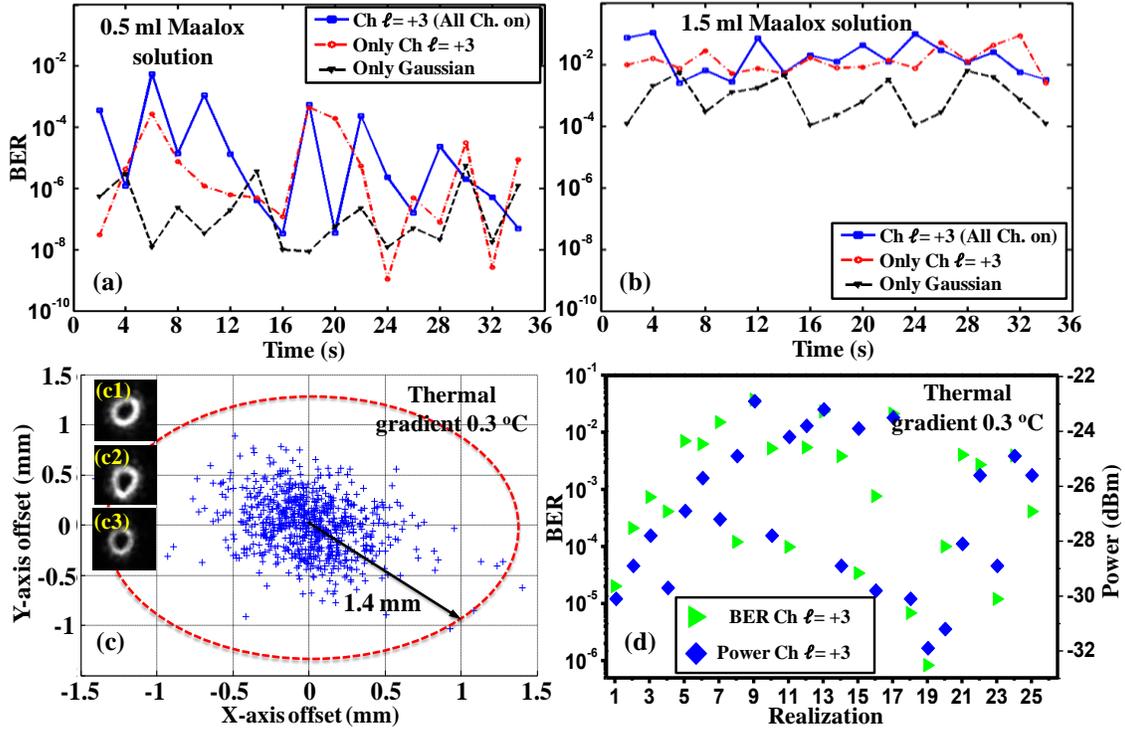

**Supplementary Figure S1** | Measured signal fluctuations and beam wander at the receiver under different scattering conditions and thermal gradients. Instantaneous BERs of channel $\ell = +3$ over 36 seconds at a fixed transmitted signal power of -26 dBm when (a) 0.5-millilitre and (b) 1.5-millilitre of Maalox solution are added into tap water (before 1-minute circulation). (c) Statistics for the displacement of the received beams with respect to the propagation axis due to thermal gradient-induced turbulence. (d) BERs and received power for the OAM $\ell = +3$ channel under various realisations of thermal gradient-induced turbulence.

**Supplementary Figure S2**

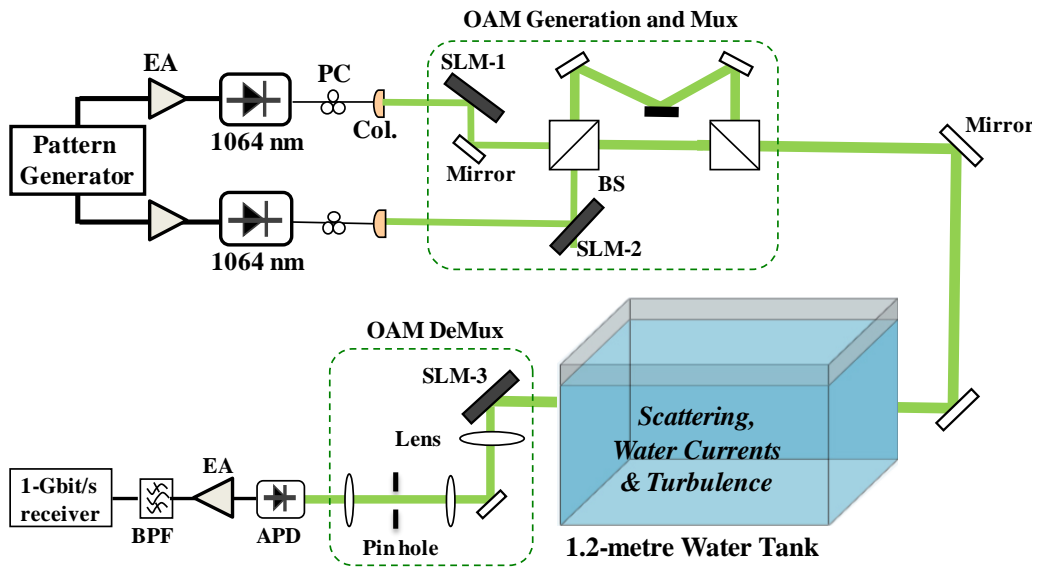

**Supplementary Figure S2 |** Experimental setup of a 4-Gbit/s OAM multiplexed underwater link using directly modulated 520-nm laser diodes. APD: avalanche photodiode. BS: beam splitter. Col.: collimator. DeMux: demultiplexing. EA: electric amplifier. Gen.: generation. LPF: low-pass filter. Mux: multiplexing. PC: polarisation controller. SLM: spatial light modulator. TIA: transimpedance amplifier.

## Supplementary Figure S3

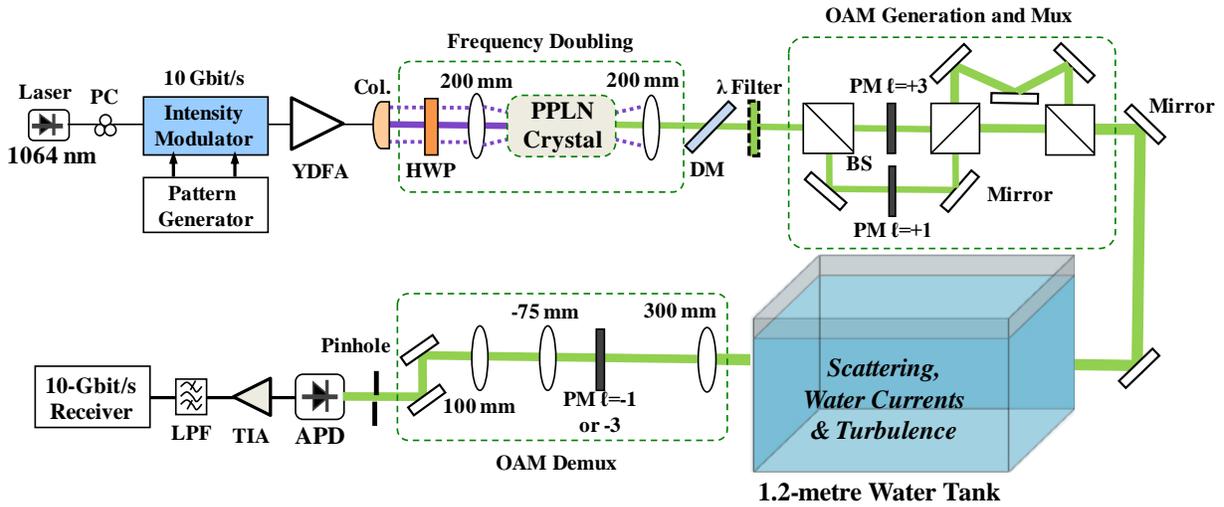

**Supplementary Figure S3 |** Experimental setup of a 40-Gbit/s OAM multiplexed underwater link using PPLN-based frequency doubling for signal generation. APD: avalanche photodetector. DM: dichroic mirror. HWP: half-wave plate. PM: metasurface phase mask. PPLN: periodically poled lithium niobate. YDFA: ytterbium-doped fibre amplifier.

**Supplementary Figure S4**

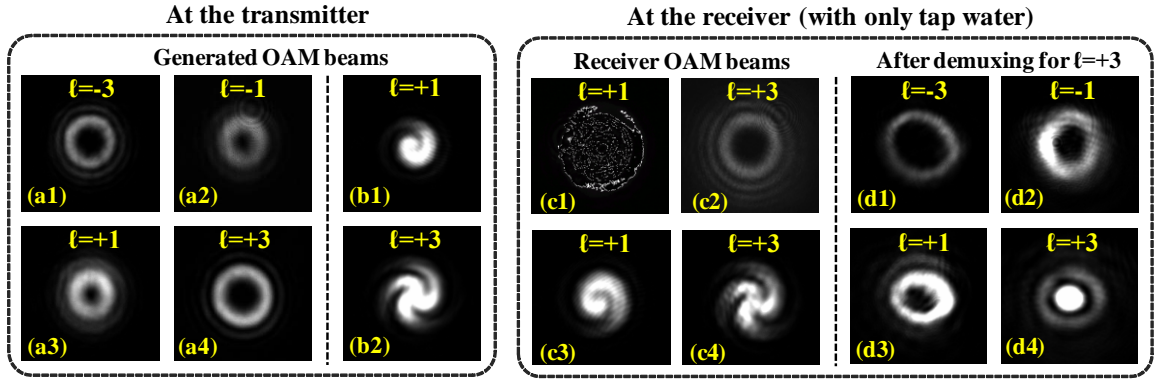

**Supplementary Figure S4 |** Captured intensity profiles at the transmitter and the receiver. (a) Intensity profiles of generated OAM beams ℓ = ±1 and ℓ = ±3. (b) Measured interferograms of generated OAM beams ℓ = +1 and +3. (c) Measured intensity profiles and interferograms of received OAM beams ℓ = +1, and ℓ = +3 after propagating through tap water. (d) Demultiplexed intensity profiles for transmitted OAM beam ℓ=+3 when a phase mask with an inverse spiral phase of ℓ = -1, +1, -3 or +3 is used at the receiver.

**Supplementary Figure S5**

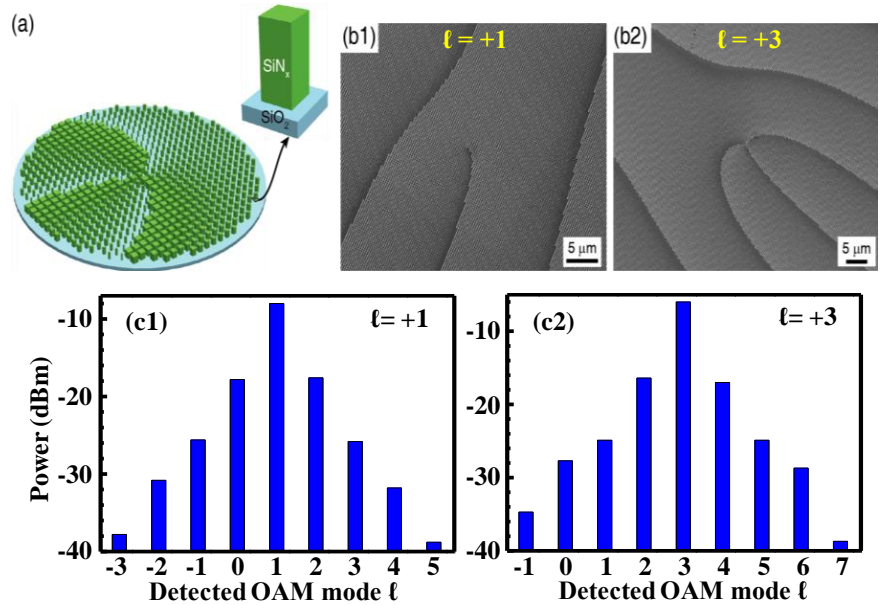

**Supplementary Figure S5 |** Dielectric metasurfaces phase mask for OAM generation and detection. (a) Schematic illustration of a blazed grating 'fork' phase mask for OAM $\ell=+3$. (Inset) The building block of the metasurface structure: square cross-section $SiN_x$ nano-posts resting on a fused silica substrate. (b) Scanning electron micrographs of the fabricated phase masks of OAM $\ell=+1$ and $\ell=+3$. (c) Measured OAM power spectrum of the generated OAM beams with $\ell=+1$ and $+3$.